# QUANTUM FISHER INFORMATION OF W AND GHZ STATE SUPERPOSITION UNDER DECOHERENCE

VOLKAN EROL
Computer Engineering Department, Okan University
Istanbul 34959, Turkey
volkan.erol@gmail.com

Abstract. We study the changes in quantum Fisher information (QFI) values for one quantum system consisting of a superposition of W and GHZ states. In a recent work [6], QFI values of this mentioned system studied. In this work, we extend this problem for the changes of QFI values in some noisy channels. We show the change in QFI depending on noise parameters. We report interesting results for different type of decoherence channels.

## 1. Introduction

The Quantum Information Theory and Quantum Computation are hot working areas that are the theoretical basis of Quantum Computers, which are described as computer technology of the future and intended to operate at very high speeds.

Quantum Fisher Information, a version of Fisher Information, developed for quantum systems, has also become a highly studied subject in recent years, as it also measures the sensitivity that systems can provide for phase sensitive tasks [1-35].

Quantum Fisher Information (QFI) is a very useful concept for analyzing situations that require phase sensitivity. This feature has attracted attention and extends the classical Fisher Information. Especially for systems with a higher QFI value, the accuracy is more clearly achieved; For example, clock synchronization [40] and quantum frequency standards [41]. Although some of the pure entangled systems may exceed the classical limit, this does not apply to all entangled systems [36]. The interaction between the quantum system and the environment not only reduces entanglement but also reduces the system's Quantum Fisher Information, in general. So we can say that researching quantum systems on QFI is important for the progress of quantum technologies. In recent studies, a single parameter, $\chi^2$ parameter, phase sensitivity was added to measure only the self-knowledge of the system under investigation [11]. Since a condition of $\chi^2 < 1$ is not provided for a general quantum system, it is understood that the system has multiple entanglement and this system provides better phase accuracy than a separable system. These quantum systems are called "useful" systems in the literature. For two-level N-particle quantumsystems, the Cramer-Rao limit is defined by the following formula [37,38]:

$$\Delta\phi_{QCB} \equiv \frac{1}{\sqrt{N_m F}} \qquad (1)$$

where $N_m$ is the number of experiments on the system being measured and $F$ is the Quantum Fisher Information value. We can write 3-dimensional vectors normalized in the nth direction of angular momentum operators, $J_n$, Pauli matrices as follows:

$$J_{\vec{n}} = \sum_{\alpha=x,y,z} \frac{1}{2} n_\alpha \sigma_\alpha \qquad (2)$$

For $J_n$, the Fisher Information of the ρ quantum system can be expressed in a symmetric matrix $C$ [10]:

$$F(\rho, J_{\vec{n}}) = \sum_{i \neq j} \frac{2(p_i - p_j)^2}{p_i + p_j} |\langle i|J_{\vec{n}}|j\rangle|^2 = \vec{n} C \vec{n}^T \qquad (3)$$

where $p_i$ and $|i\rangle$ represent the eigenvalues and eigenvectors of the $\rho$ system, respectively, and the matrix $C$ is defined as

$$C_{kl} = \sum_{i \neq j} \frac{(p_i - p_j)^2}{p_i + p_j} [\langle i|J_k|j\rangle\langle j|J_i|i\rangle + \langle i|J_i|j\rangle\langle j|J_k|i\rangle] \qquad (4)$$

The largest $F$ value between the $N$ options is selected and averaged over $N$ particles. The Fisher Information value is calculated as the greatest eigenvalue of the $C$ matrix. This definition is expressed by the equation:

$$\bar{F}_{max} = \frac{1}{N} \max_{\vec{n}} F(\rho, J_{\vec{n}}) = \frac{\lambda_{max}}{N} . \qquad (5)$$

## 2. Studied Quantum System and Reported Results

In this section, we introduce three qubit systems in decoherence channels. Particularly the system studied is superposition of a W and a GHZ state [6],

$$|\varphi^N\rangle = \alpha |W^N\rangle + \sqrt{1-\alpha^2}|GHZ^N\rangle \qquad (6)$$

Here $N$ is the number of qubits, and α is the superposition coefficient. In this work, we consider the case for $N=3$. The decoherence channels are *amplitude damping, phase damping* and *depolarizing* respectively. Then we study the QFI of the three qubit system for different scenarios. As a first scenario, we take decoherence parameter $p$ as 0.1 and $α$ as the variable of the system. In Figure 1, changes in QFI values for three channels are showed for $p = 0.1$ constant value.

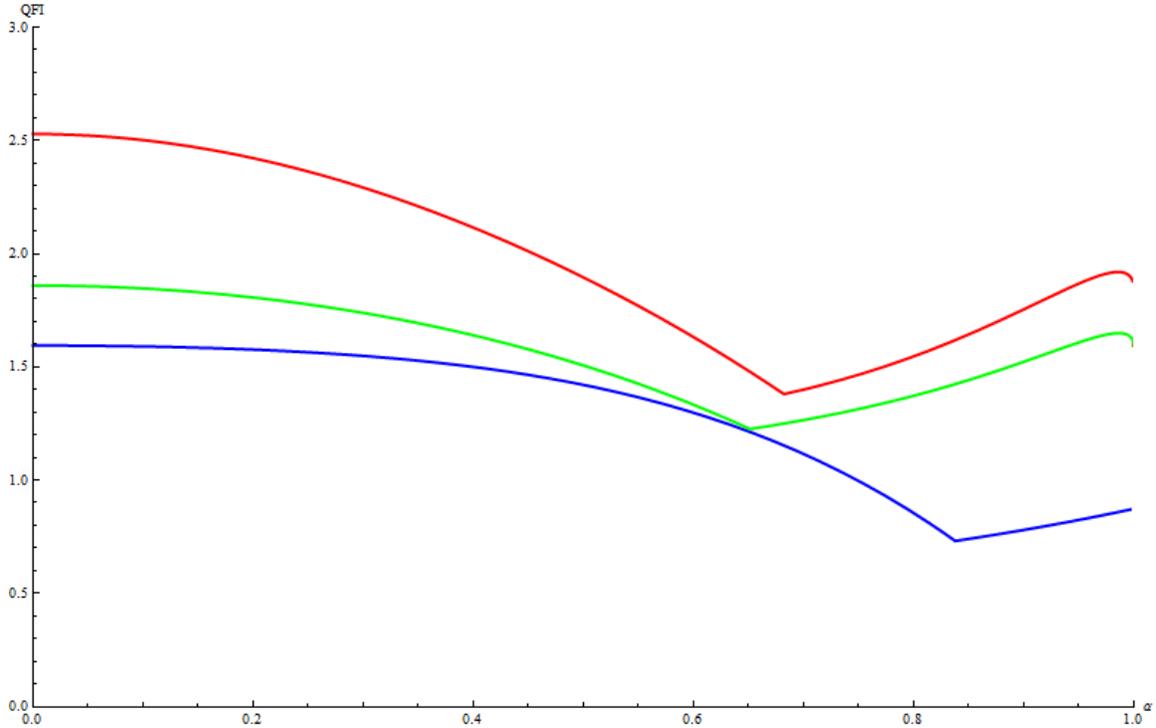

Fig. 1. Changes in QFI for decoherence channels for *p = 0.1* (Red: Amplitude Damping, Blue: Phase Damping, Green: Depolarizing)

In the second scenario, we find the QFI as functions of $\alpha$ and $p$. We take values of $\alpha$ and $p$ between 0 and 1.

In Figure 2, we show in contour plot, the changes of QFI values in three channels.

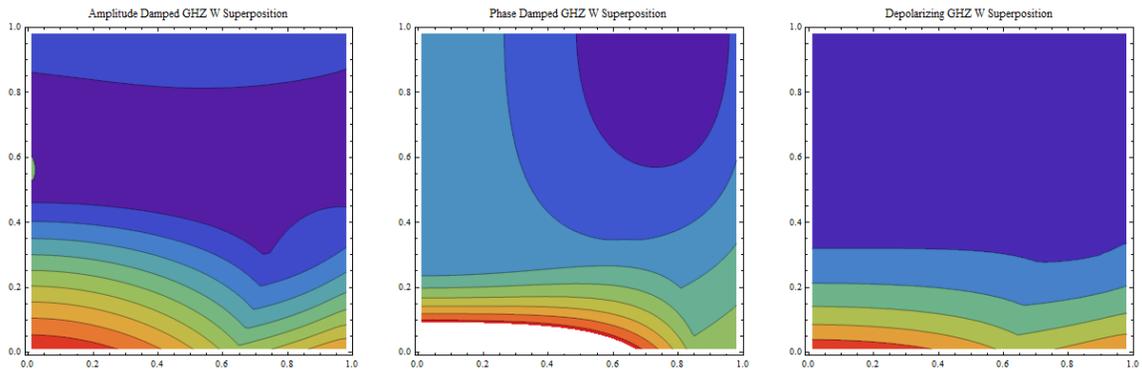

Fig. 2. QFI Values as a function of $\alpha$ and $p$ (contour plot)

In Figure 3, we show in 3D Plots the changes in QFI values.

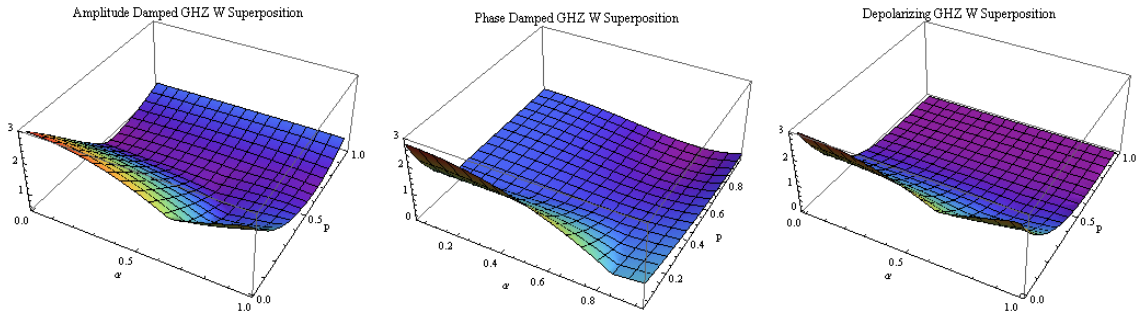

Fig. 3. QFI values as function of α and p (3D plot)

We have some observations about the results: firstly, regardless the value of the noise in the channel, dependence to α superposition coefficient effects the QFI values dramatically. In amplitude damping and depolarizing channel cases, it was observed that until to some values of α (~0.65), QFI value decreases and after this point, it starts increasing. In these cases, it behaves like a local minimum. This situation is not similar for phase damping case.

Second observation is that this superposition state is more resistant to phase damping channel. It can be viewed from Figures 2 and 3 (purple areas are smaller).

It was shown in [6] that this superposition is "useful" unless α is around 0.75 value.

## 3. Conclusion

We studied the changes in QFI values for a quantum system which is a superposition of $W_3$ and $GHZ_3$ states. In this work, we extended this problem for the changes of QFI values in amplitude damping, phase damping and depolarizing channels. We showed the changes in QFI depending on noise parameters and superposition coefficient. We reported interesting results like the dramatical decreases of QFI values in amplitude damping and depolarizing channel cases. One of main observations arrived is that this nearly-useful state is more resistant to phase damping channel noise comparing to other decoherence channels.